# Insights on structure and influence from the adjacency and Laplacian eigenspectra of intersecting ring networks

Agathe Bouis*, Ruaridh A. Clark, Malcolm Macdonald

Applied Space Technology Laboratory, Department of Electronic and Electrical Engineering, University of Strathclyde, Glasgow, UK.

## ABSTRACT

A network's community structure commonly impacts its functions. For instance, networks seeking synchronisation will see this process follow the topology's hierarchical and community structuring. Herein, the interplay of network adjacency and Laplacian eigenspectra is shown to uncover hierarchical influence and community structure. Ring networks embedded with hubs of high connectivity are first analysed to characterise the differing insights of the adjacency and Laplacian eigenspectra. This understanding is then transferred to the case study of satellite networks composed of intersecting rings. From these scenarios, it is found that where the adjacency reflects a network's structure, the Laplacian detects nodes' influence profiles. The adjacency identifies the number and relative sizes of salient network structures, for example hubs of high connectivity. In contrast, the Laplacian decomposes these structures in sets of nodes with equivalent influence profiles, where influence emerges due to variations in nodes' connectivity.

## I. INTRODUCTION

Networks represented by graph matrices can be used to study real world systems, where a graph consists of an ensemble of node (vertices) with connections between each other (edges) [1]. Node connections are typically not random, instead the distribution of the number of connections per node, or node degree, displays high levels of ordering and hierarchical organisation relating to the system's functions [1, 2]. This underlying inhomogeneity in network connectivity associated with patterns in edge density at both a global and local scale is called community structure. Communities, or clusters, are groups of vertices which often play similar roles within the system. They are distinguished by their high connectivity density within the group and only sparse connections outwith [3, 4]. Communities can develop at both global and local levels and are distinguished by the reach of their nodes. It is important to note that most real world networks typically consist of interlocked, interacting, nested communities acting at different scales [2]. Herein, these community patterns are studied through two properties: *structure* and *influence*.

Structure is a topological property used herein to describe the principal community structures which emerge because of the dominance and global reach of certain clusters over the network. In contrast, influence is a measure of how strongly a node's or community's dynamical state affects the collective behaviour of the network (e.g. synchronisation) [5, 6]. Influence, as defined by [5] in the wider context of communities of dynamical influence, accounts for the interplay between network topology and its dynamics [6], where topology describes the arrangement of the network elements (edges and nodes). Influence stems not only from direct connections and local reach, but also indirect connections, accounting for nodes' effective leadership over neighbouring nodes' states. That is, it relates to nodes' ability to propagate information effectively. Herein, the term community of dynamical influence is used to describe the set of nodes with the same influence profile. Structure and dynamical influence are two of many approaches to analyse a network's topology.

* Contact author: agathe.bouis@strath.ac.uk

Community detection can be performed by spectral analysis [7-9]. This consists of studying a graph's representative matrix $M$, typically its adjacency or Laplacian matrix, and examining its eigenvalues and associated eigenvectors. As derived from the Rayleigh quotient [10, 11], a network's largest eigenvalue(s) takes on values between the network's maximum degree (found at the hubs) and its average degree. The largest eigenvalue, $\lambda_1^M$, also known as the network's spectral radius, is known to be linked to information dissemination in networks [12]. In epidemic propagation models, the steady-state infection is determined by an epidemic threshold $t_c = 1/\lambda_1^M$. If the effective infection is lower than this threshold, the network is virus free [13]. Similarly, the synchronisation transition from incoherence to coherence in networks of coupled phase oscillators is related to the network's coupling strength $t_c$ [14]. The eigenvectors associated with this first eigenvalue have also been used to determine network's most central nodes, either in terms of their information/disease spreading capabilities with the Laplacian [6], or on the basis of which nodes are visited most often by a random walker with the adjacency matrix [15].

Eigengaps, that is the difference between adjacent eigenvalues, are known to give insights into the detection community structure, with the largest eigengap being noted to typically indicate the number of distinct community present in the network topology [8]. Knowledge of the largest eigengaps can also be used to inform eigenvector-based clustering methods. For these spectral methods, a set of $p$ eigenvectors of the matrix $M$ are computed. In $p$-dimensional eigenvector space, communities appear as groups of points separated from each other [5, 16]. They can thus be detected using clustering algorithms such as k-means [17]. Eigenvector clustering has been found to successfully detect graph partitions, most notably for communities with no overlap, meaning that no node belongs to more than a single community [2]. Eigenvector decomposition and clustering helps overcome the challenge of nodes' and edges' lack of canonical ordering and/or absent positional data [18]. Compared to the more commonly used statistical methods such as degree distribution, average degree, centrality, and modularity, spectral analyses have proven to be more effective at determining graph structural characteristics [19, 20]. While spectral analyses offer the potential to identify groups of interacting vertices given a network's underlying topology [21, 22], insights are dependent on the choice of representation matrix, most commonly its adjacency or Laplacian matrix [8]. Although the Laplacian is often the matrix of choice for community detection purposes [23-25], especially when looking at the discovery of influential nodes due to the matrix's connection to consensus dynamics [5], many works have analysed the adjacency spectrum with the same goal [6, 15]. Despite the commonality of this dual usage of the adjacency and Laplacian matrices, to the best of the authors' knowledge, no clear intuitions have been established regarding what complementary insights the adjacency and Laplacian might offer.

Herein, spectral analyses are performed for ring-based networks. Ring networks – also known as circulant graphs [26], cyclic graphs [27], distributed loop networks [28-30], chordal rings [31], or multiple fixed step graphs [32, 33] – are of high interest to mathematics and engineering due to their simplicity, expandability, and regularity [28]. Ring networks represent a special case of weak circulant network coupling where each node is connected to its $\delta_R$ nearest neighbours [26, 34]. These topologies appear deceptively simple and yet, as they form closed loops of signal transmission with a signal being able to cycle across the network to come back to its node of origin, their synchronisation patterns have proven to be dynamically complex [35].

In a network, the process of synchronisation consists of the evolution of node states over time due to the nodes' interactions. Synchronisation can be described as stable/unstable, indicating the process' final state's stability, but also complete/partial, dependent on the number of nodes in agreement with each other at a given time. Synchronisation is heavily dependent on a network's topology and its hierarchical scales [36], with the process typically developing at a local community level and building to the global [22]. Ring networks, which lack any topological scale or community structure, can see the emergence of stable or unstable synchronisation, partial synchronisation, or even nonlinear chimeras [37, 38]. Synchronisation was found to be highly dependent upon coupling strength, with a threshold of around 68% being necessary to ensure complete synchronisation [39, 40]. Ring networks also display characteristic and distinct eigenspectra. This is of particular interest given the application of ring networks in the domains of telecommunications, computer science, engineering (see [34] and references therein), This paper focuses on Earth-orbiting satellite systems which heavily feature ring networks.

Modern satellites are increasingly being equipped with Inter-Satellite Link (ISL) transmitters which allow them to communicate when in range of each other [41], meaning that the satellites are close enough to transmit and share information. Such capabilities allow satellite communication systems to be abstracted as time-varying networks [42], where each network edge represents a temporary communication opportunity between two satellites. Among the most ubiquitous topologies of satellite networks are Walker-delta constellations [43]. These constellations consist of satellites placed in circular orbits (around the Earth) at the same altitude and orbital inclination but distributed across in longitude over the Earth [44, 45]. As a result, Walker-Delta form intersecting rings set in complementary orbital planes [46]

Intersecting ring networks and thus Walker-delta satellite constellations combine the properties of rings, including their lack of topological scales and low connectivity, with hubs of high connectivity at the rings' intersections. These characteristic features are expected to provide differing adjacency and Laplacian eigenspectra [47]. This topology also brings together two complex chaotic phenomena: incoherent ring synchronisation and cluster synchronisation at the rings' intersections. Satellite systems, much like many engineered networks, require global synchronisation to secure their operations; a challenging requirement given the prevalence of ring network in satellite systems and their complex synchronisation. New research is looking at the potential of decentralised synchronisation driven by local node (satellite) interactions [48, 49]. Although arguably more complex than the Kuramoto model's mean field regime [50], the dynamics of satellite time synchronisation are similarly linked to a network's topology [51]. Uncovering the topology of intersecting ring networks, most notably the interplay of the structure and dynamic influence within them is key to better understanding these networks' patterns of synchronisation and in the future, engineer solutions to ensure their global synchronisation. Herein, intersecting rings are thus studied through their adjacency and Laplacian eigenspectra and their differing spectral insights are compared.

# II. METHODOLOGY

The topology of an undirected network, $G = (E, \mathcal{E}, A)$, can be described with a finite number of nodes (agents) $E = \{1, 2, \ldots, n\}$ and a set of edges $\mathcal{E} \subseteq E \times E$. Each edge $(j, i) = (i, j) \in \mathcal{E}$ represents a link between the node pair $(j, i)$, such that communication between the nodes is enabled. The existence of edges is capture in the adjacency matrix $A = [a_{ij}]_{n\times n}$, where $a_{ij} = 1$ denotes an unweighted edge. A weighted edge is considered as $a_{ij} > 0.1$. A node $i$'s degree is indicated as $\delta^i$. Adapted from the adjacency matrix, the standard Laplacian matrix is defined as $L = (D - A)$, where $D$ is a diagonal matrix whose diagonal elements $D_{ii} = \delta^i$ are the degree of the node $i$. Eigenvectors are vectors whose direction remains unchanged when subjected to a linear transformation. Instead, the vectors are scaled by a constant factor $\lambda$ (eigenvalue) when the transformation, $\lambda\mathbf{v} = \mathbf{v}M$, is applied. $M$ is a representation matrix and is used as a stand in for either the adjacency, $A$, or the Laplacian, $L$.

## A. Community Detection

A network's spectrum is herein defined as the sorted eigenvalues of the adjacency and Laplacian matrices, giving the adjacency spectrum and Laplacian spectrum, respectively. These eigenvalues are sorted in descending order according to their Real parts, that is,

$$\mathbb{R}(\lambda_1^M) \geq \mathbb{R}(\lambda_2^M) \geq \cdots \geq \mathbb{R}(\lambda_n^M). \qquad (1)$$

The eigenvalues' associated eigenvectors are sorted in a similar fashion, resulting in an eigenvector matrix, $V_{n\times n}^M = [\mathbf{v}_1^{\mathrm{M}}, \ldots, \mathbf{v}_n^{\mathrm{M}}]$.

Comparisons of the adjacency and Laplacian spectra have noted that the larger the degree extreme difference, that is the difference between the maximum and minimum degree, the more differentiated the eigenvalues and eigenvectors will be [47]. As such, for a d-regular graph, representation matrices are simply related by an affine transformation, while for extreme degree differences, spectral clustering results will differ [47].

Eigengaps are calculated for both the adjacency and Laplacian as the difference between adjacent eigenvalues (such that herein all eigengaps have negative values),

$$\mathrm{g}_{12}^{M} = \lambda_2^M - \lambda_1^M. \quad (2)$$

Eigengaps are known to reflect community clustering in networks [8]. More specifically, the index of the largest eigengap $\mathrm{idx}(g_{\max}^M)$ is understood to indicate the most dominant partitioning, meaning the number of sets of nodes (subgraphs) into which it is easiest to split the networks by removing edges connecting the communities.

Further analysis of the network properties can be conducted by examining a matrix's eigenvectors and deriving its associated local eigenvector centrality measures [52]. Eigenvectors are useful for community detection: for a given eigenvalue, a node's eigenvector magnitude relates to its influence as the network is being pulled in as many directions as the ranks of the eigenvalue, $\mathrm{idx}(g^M)$. Thus, for the eigenvalue at rank $i = 3$, the associated eigenvector gives insights into nodes' influence as the network is pulled in three directions. Nodes with the highest eigenvector magnitude across the network's most significant eigenvalues are the most influential. Given knowledge of a canonical ordering of nodes' and/or of their positional data, a heatmap of eigenvectors can be plotted and analysed for community clusters, where nodes part of the same community will typically be influential. If this data is absent or no clear ordering exists, a local eigenvector centrality measure is calculated, capturing the influence of nodes within their communities. The choice of matrix, the adjacency or the Laplacian, will define whether that influence spans global or local scales. Local eigenvector centrality [52], $c(i)$, is defined for each node $i$ as the Euclidian norm of a truncated eigenvector matrix, $T^M$ , corresponding to the eigenvector series of interest,

$$T_{n\times(y-x)}^M = \left[\mathbf{v}_x^M, \mathbf{v}_{x+1}^M, \dots, \mathbf{v}_y^M\right], \quad (3)$$

$$c(i) = \left[\sum_{k=x}^{y} \left(\mathbf{v}_k^M(i)\right)^2\right]^{\frac{1}{2}}, \quad (4)$$

where $x$ and $y$ ( $(x, y) \in [1, n]$ and $y > x\}$ ) ranks of the first and last columns selected from the sorted eigenvector matrix, $V^M$. The new matrix contains $n$ rows (for each of the nodes) and $(y - x)$ columns (for each of the selected eigenvectors). Each node's local eigenvector centrality, $c(i)$, can take on values between $[0, 1]$. Local eigenvector centrality considers the most topologically relevant scale for detecting node importance in reference to the choice of the truncated matrix, reflecting localised, community-considerate assessments of centrality. As such, nodes with the most influence take on values closer to 1 as they are influential across many of the eigenvector's columns. Nodes with the least influence take on values tending toward 0.

## B. Ring Eigenspectrum

Ring networks are special cases of circulant matrices. Circulant matrices are defined square matrices where all rows are composed of the same elements and each row consists of the elements of the previous row shifted to the right of the preceding row [26]. For ring networks, nodes are only connected to $\delta_R$ of their sequential neighbours ( $\delta_R/2$ nodes on each "side" of the node), while the cyclic permutation is limited to one rightwards column shift. Thus, for a ring of connectivity $\delta_R = 2$, the adjacency matrix takes on the following form,

$$A_R = \begin{bmatrix} 0 & 1 & 0 & \cdots & 0 & 1 \\ 1 & 0 & 1 & & 0 & 0 \\ & \vdots & & \ddots & \vdots & \\ 0 & 0 & 0 & \cdots & 0 & 1 \\ 1 & 0 & 0 & & 1 & 0 \end{bmatrix}. \quad (5)$$

It is important to understand the seemingly simple spectra of single ring networks before delving into the more complex networks motivating this research. Graph data can be analysed using common signal processing techniques: a graph signal is a vector of data, with each entry corresponding to a node data point. Graph frequencies are evaluated based on how node data change between nodes. The underlying network thus defines how the frequencies are calculated, with changes in topology impacting local variations [53]. In turn, these frequencies can be analysed using signal processing methods. Rings networks, with their circulant matrix where each row is a circular shift from the previous row [54], act as a circular convolution operator to any signal implemented upon them, with the eigenvectors of the network being, in fact, the columns of the Discrete Fourier Transform (DFT). Through this process, the signal, meaning the nodes' data, is effectively viewed as one period of length $n$ (number of nodes) [53]. Much like the columns of the DFT matrix representing equal partitions of the unit circle [53], the eigenvectors of the network's Laplacian take on the values of the primitive root of unity [55],

$$\omega^M(i) = \exp\left(\frac{2\pi i}{n}\right). \quad (6)$$

$$\mathbf{v}^M(i) = \frac{1}{\sqrt{n}}\left(1, \omega^j, \omega^{2j}, \dots, \omega^{(n-1)j}\right)^T, \quad j = 0, 1, \dots, n-1. \quad (7)$$

Where $\omega$ is the primitive $n$-the root of unity and $i$ is the imaginary unit.

As a result, for real-symmetric circulant matrices, pairs of consecutive eigenvalues have equal magnitudes, except for the first dominant eigenvalue, with the real and imaginary components of the primitive roots of unity splitting across two sets of eigenvectors. More practically speaking, this results in the network's eigengap spectrum displaying stepped behaviour as eigenvalues one and two, three and four, and so forth, are identical. The adjacency spectra and eigengap spectra of a 100-node ring's spectrum with different connectivity levels, $\delta_R = \{2, 4, 6\}$ are shown in Fig. 1.

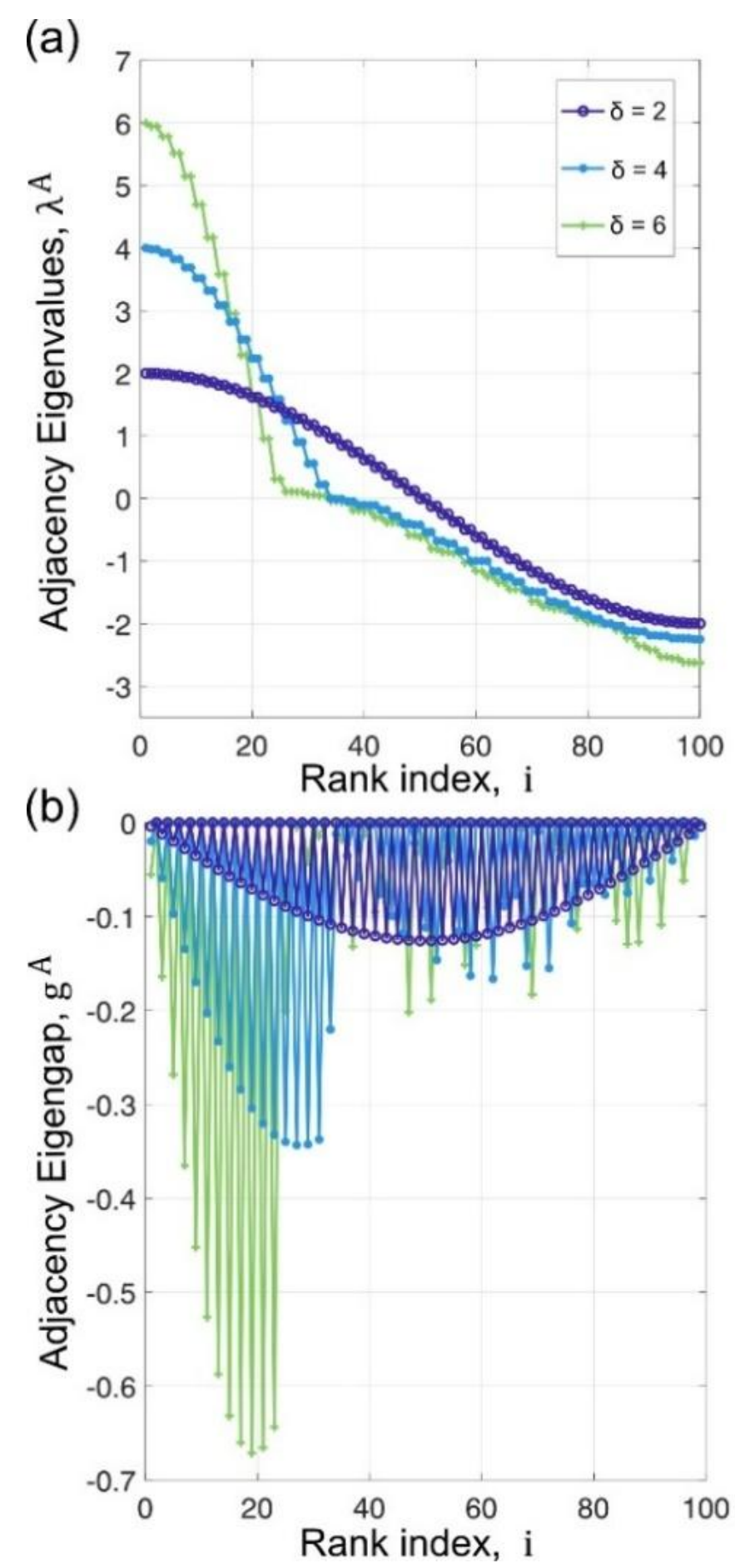

Fig. 1: (a) Adjacency and (b) adjacency eigengap spectra of 100-node ring with $\delta_R = \{2,4,6\}$.

Node connectivity across the rings is symmetrical, with nodes connecting to $(\delta_R/2)$ nodes on either of their sides. Note that no clear community structure is observed (Fig. 1), with no single eigengap dominating the spectra.

This distinct eigenspectrum – fingerprint – makes ring networks a useful foundation upon which to study community formation as it can be dismissed from the studied networks' own eigenspectra.

## C. Simulation Setup

To uncover the insights and interplay of the adjacency and Laplacian spectra, two sets of analyses are performed. The first set investigates aggregated ring networks consisting of a simple ring of fixed connectivity embedded with clusters of highly connected nodes. The second set of analyses are performed over satellite networks consisting of intersecting rings. Clusters of high connectivity emerge at the rings' intersection.

### *1. Aggregated Rings*

The aggregated ring networks are obtained by taking a simple ring (Fig. 1) of $n$ node and connectivity, $\delta_R$, from which $k$ sets of adjacent of nodes are selected. The sets are denoted $h_k$, where $|h_k|$ is the cardinality of the set, meaning the number of nodes belonging to the set. Selected nodes are connected to all other nodes within their sets, resulting in clusters (hubs) of high connectivity embedded on the ring topology. Analyses are performed for four scenarios, $S_k$, with a maximum hub count, $c$. Simulations are defined as,

$$S_k = \{(h_k)_{k=1}^{c} : |h_k| \in \mathbb{Z}^{0+}\}, \tag{8}$$

The size of hubs, $|h_k|$, and maximum hub count, c, are specified by the analysis scenario parameters.

Hubs are embedded on the ring to be as equidistant from each other as possible. This means that the number of nodes between the hubs is set to be the same (or as close to) as possible. The hubs are placed at IDs spanning $D_k^{start}$ to $D_k^{end}$ for each hub $k$, where D can take on a range matching $E = \{1, 2, \dots, n\}$,

$$u = round\left(\frac{n}{c}\right), \tag{9}$$

$$\begin{aligned} D_k^{start} &= (k-1)_{k=1}^{c} \cdot u + 1, \\ D_k^{end} &= (k-1)_{k=1}^{c} \cdot u + |h_k|. \end{aligned} \tag{10}$$

Note that given the size and number of the hubs in each analysis, there might not be the exact same number of nodes between each hub.

A feature of the artificial construction of the hubs is that two types of nodes are distinguished for each hub: "central" and "gateway" nodes (Fig. 2).

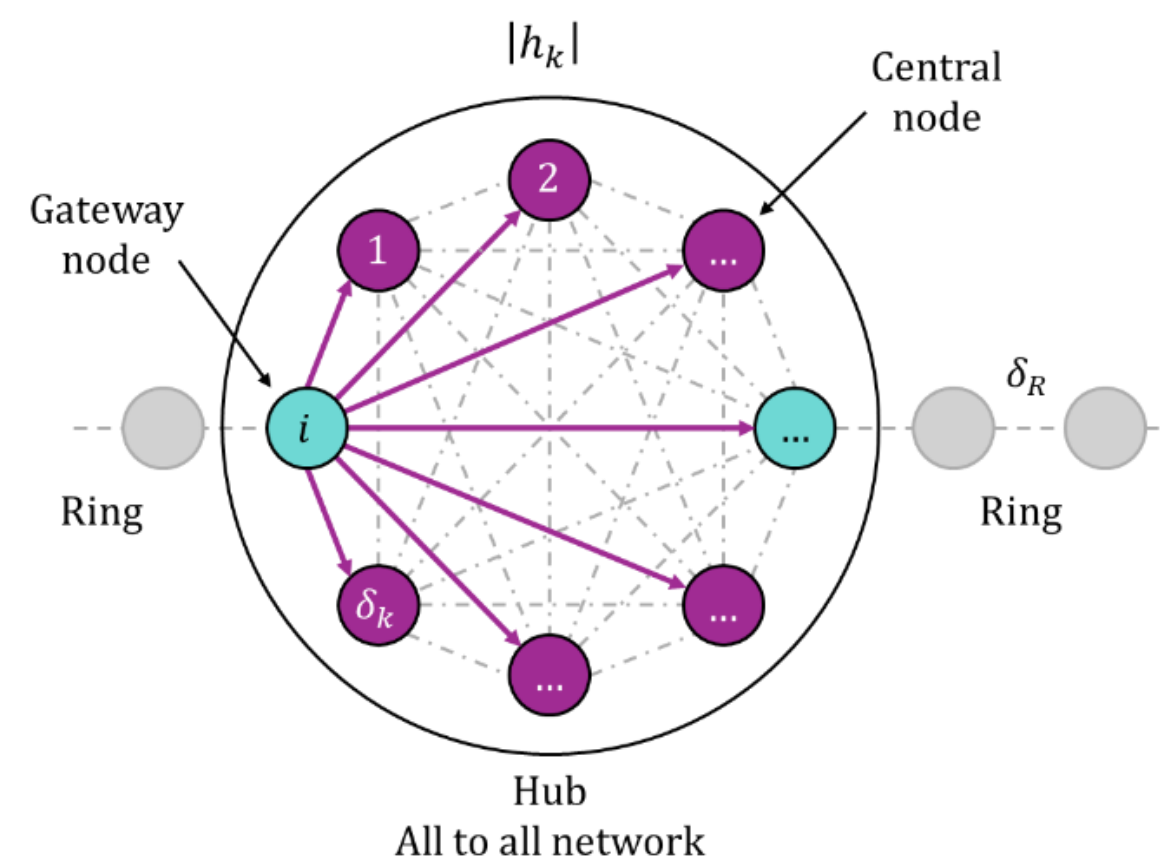

Fig. 2: Hub Structure

Central nodes are those nodes which only connect to other hub nodes, while gateway nodes connect beyond the hub. As such, central nodes have a degree equal to the number of other nodes present within their hub $k$ ($|h_k| - 1 = \delta_k$). In contrast, gateway nodes have a higher connectivity (degree) as they connect to central nodes and ring-nodes, acting as gateways to the hub and having influence over both the hub nodes and the ring-nodes. The gateway nodes' exact degree depends on where they are on the hub border, with nodes most to the edge of the hub having the highest degree as they are connected to all other hub nodes as well as to $\delta_R/2$ ring nodes for a total of $\delta = |h_k| - 1 + \delta_R/2$. Note that as the hubs are symmetrical, gateway nodes come in pairs with both nodes in the pair having the same degree.

### 2. Intersecting Rings

The intersecting rings networks are extracted from satellite communication networks. Satellites equipped with Inter-Satellite Link (ISL) sub-systems can communicate when in view and in range of each other. Satellites constellations can be abstracted as time-varying networks where satellites are nodes and edges represent the contact opportunities between satellites which appear and disappear as the satellites move with respect to each other in orbit around the Earth. Ring networks are the basis of many satellite constellations given the underlying principles of celestial mechanics, i.e. satellites orbit in circles (rings) around a gravitational point (e.g. the Earth). Herein, attention is paid only to the engineered ring networks which form the basis of many constellations [43, 46, 49]. These networks are modelled by locating satellites at equal angular intervals on a ring, that $360°/m$, where $m$ is the number of satellites in the ring. For a fixed ISL range, the denser the ring ($m$ per ring), the higher the nodes' degree connectivity, $\delta_R$.

Walker-Delta constellations, that is constellations of intersecting rings, are a common solution to maximise the geometric coverage of the Earth while minimising the number of satellites to be placed in orbit. The Walker-Delta network of interest contains a total of $n$ satellites, with $m$ satellites evenly spaced in each of $P$ orbital plane, so that $n = m \cdot P$ (where the $m$ satellites form a ring network of connectivity $\delta_R$ ). The $P$ orbital planes are equally spaced in 3-dimensional space [56] (see [57] and [58] for further detail regarding celestial dynamics and descriptions of orbital parameters), with the $P$ rings all having a common focus, that is a common centre point: the Earth, as seen in Fig. 3.

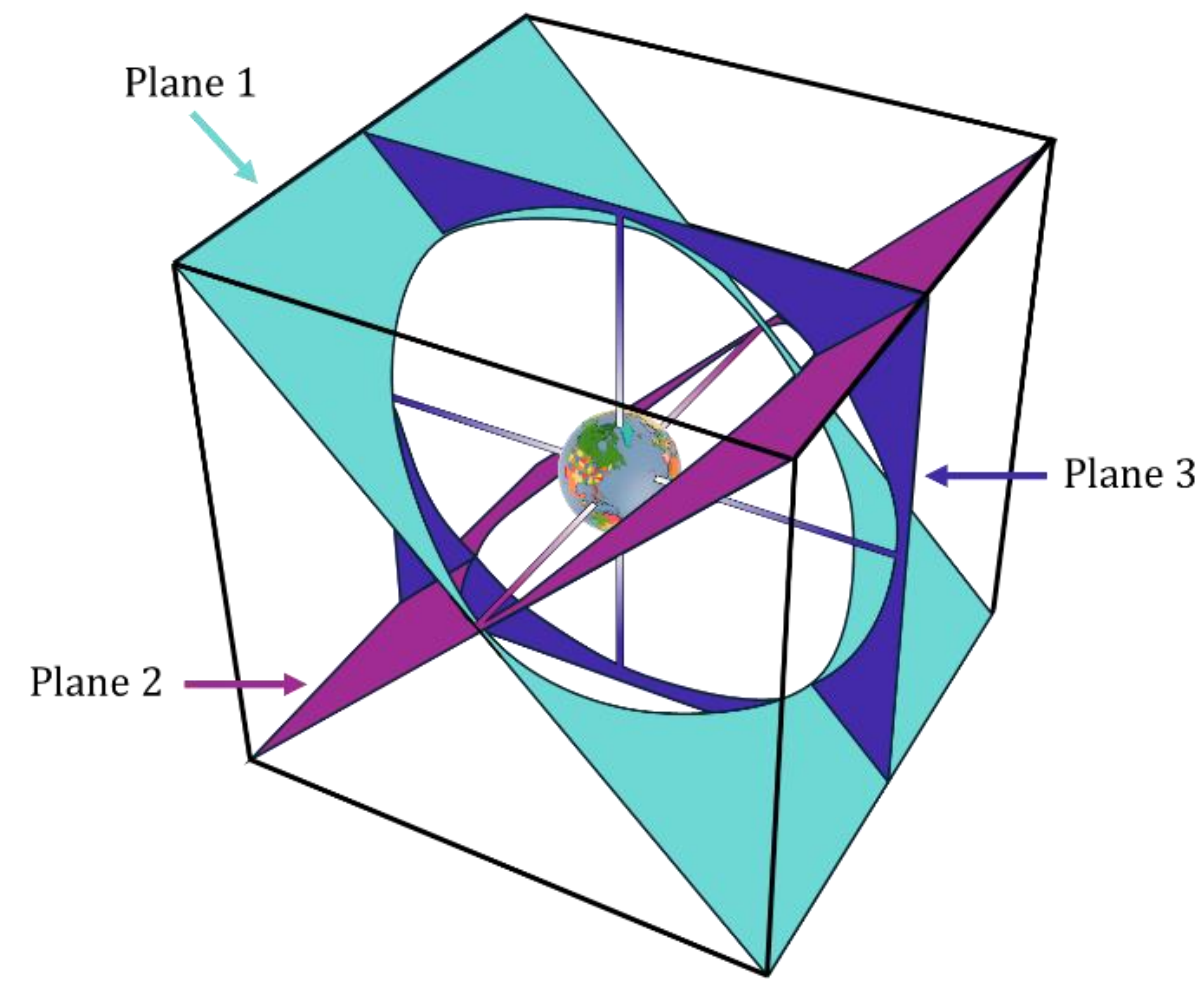

Fig. 3: Constellation set-up for a three-ring network. For a two-ring network, only Planes 1 and 2 are considered.

An interplane phasing $F$ is used to define the number of "empty slots" between the first satellites in each orbit plane. With a phasing degree, satellites are not offset and would thus collide in their respective rings. Herein, this phasing is fixed to one "slot", where the angular distance represented of that slot depends on the number of satellites on this plane's orbital ring, $360/m$ . Herein, constellations are modelled using the "sdp4" orbit propagator for a period of five hours with a step size of ten seconds.

When two rings intersect, hubs are naturally created at their crossing point. This is because satellites come in range of each other's communication payloads. Consequently, satellites can communicate with each other, resulting in the creation of network edges between the nodes. Node connectivity therefore increases smoothly starting from the hubs' edges until their cores, peaking to a maximum near the rings' intersections such that central nodes rather than gateway nodes are the most well-connected. This is a reversal from what was observed with the aggregated rings graphs. This pattern means that, unlike the aggregated ring, there are no sudden discontinuities in node connectivity distribution, nor are there clear demarcations between gateway and core communities.

As a result of the network's dynamics (satellite motion), the hubs created at the rings' intersections are not symmetric: nodes on different side of the hubs are likely to have different degrees. To mitigate against the effect of this asymmetry, where instantaneous connections may not reflect the actual network structure, thereby creating noisy results, analyses are instead performed over time-averaged adjacency and Laplacian matrices,

$$\bar{M} = \frac{1}{tw} \cdot \sum_{\tau=1}^{tw} M(t+\tau), \quad \bar{M}(\bar{M} < 0.1) = 0, \tag{11}$$

where the $tw$ are time windows comprising multiple time step $\tau$. These matrices are calculated as the mean of the time-dependent connections. Herein, $tw = 10$ is used as it has been found to offer a balance against the effect of instantaneous connections, while limiting the extent of the averaging such that the networks' structure remains comprehensible (see Appendix A for details regarding the selection of the averaging window).

# III. RESULTS

Aggregated ring networks are used as toy graphs to gain insights into the structure and influence gained from adjacency and Laplacian matrices. The observations built therefrom are applied to the case study of intersecting ring networks.

## A. Structure

By adding salient structures to a simple ring network, it is found that the adjacency eigenspectrum reflects graph structure. In Fig. 4, hubs – consisting of sets of fully interconnected nodes – are embedded onto a series of rings, and their sizes are varied to show the associated shifts in eigenspectrum. The adjacency spectrum identifies the numbers and relative sizes of hubs. The four simulations are defined as,

$$S_1 = \left\{ (h_k)_{k=1}^{c=[1,4]} : |h_k| = 10 \right\}, \tag{12}$$

$$S_2 = \left\{ (h_k)_{k=1}^{c=[1,4]} : |h_k| = \{10, 11, 12, 13\} \right\}, \tag{13}$$

$$S_3 = \left\{ (h_k)_{k=1}^{c=[1,4]} : |h_k| = \{10, 16, 17, 18\} \right\}, \tag{14}$$

$$S_4 = \left\{ (h_k)_{k=1}^{c=[1,4]} : |h_k| = \{20, 10, 11, 12\} \right\}. \tag{15}$$

The magnitude of the adjacency's eigenvalues is associated to the embedded hubs' sizes. As observed in Fig. 4, it appears that each hub is associated to one eigenvalue, $\lambda_k^A$, with that eigenvalue's magnitude depending on that hub's connectivity, $\delta_k$, (i.e. its size), keeping in line with the known lemma: $\mathrm{mean}(\delta_k) \leq \lambda_k^A \leq \max(\delta_k)$ [59]. This is likely due to the limited connections between the hubs as hubs are connected via ring arcs with a degree $\delta_R$. As a result, the higher the hub connectivity (size), the larger the magnitude of the associated eigenvalue. In turn, the larger the hub's eigenvalue's magnitude, the more it will be differentiated from the ring's eigenvalues. This is supported by comparing Fig. 4 (a.i) and Fig. 4 (b.i): therein the gap, which shows the difference between the hub's and ring's eigenvalues, is larger for the 20-node hub than the one of the smaller 10-node hub.

The adjacency's non-zero eigengaps reflects the relative sizes of the hubs embedded on the ring. Eigenvalues are sorted from largest to smallest according to their real parts (Eq. (1)). Thus, in the case where all hubs have the same connectivity (Fig. 4 (a)), all the eigenvalues associated with the hubs ($\lambda_1^A, \ldots, \lambda_c^A$) have the same value. This results in a single non-zero eigengap at $i = c$ as there is no difference between the hubs' eigenvalues. This gap's magnitude, $|g_c^A|$, stays constant as the number of hubs increases (Fig. 4 (a.i) to (a.iv)) as adding more hubs does not affect the hub nodes' connectivity, $\delta_k$, (Fig. 2), and therefore does not change their eigenvalues.

If hubs have different sizes, their eigenvalues will differ, resulting in all eigengaps up to the rank $i = c$ being non-zero, with the largest gap indicating the set of the most dominant structures (hubs) in the network (Fig. 4 (b), (c), and (d)). Keeping in mind that eigenvalues are sorted according to their magnitude, these non-zero gaps indicate the difference in hub sizes. Taking as example the case of Fig. 4 (d.ii), the gap at $i = 1$ is difference between $\lambda_1^A$ (associated with the largest 20-node hub) and $\lambda_2^A$ (associated with the 10-node hub). The second gap at $i = 2$ corresponds to the difference between the 10-node hub and the ring. Note that this last gap has the same magnitude as the gaps in Fig. 4 (a). When adding more hubs the simulation such as in Fig. 4 (d.iv), the first gap is now the difference between the 20-node ($\lambda_1^A$) and 12-node ($\lambda_2^A$) hubs' magnitudes. As a result, the eigengap's magnitude shrinks from Fig. 4 (d.ii). The second gap's magnitude is now much smaller as it only accounts for the difference between the 12-node and 11-node hub. Note again that the last gap's magnitude at $i = c$ remains fixed as it is still the difference between the 10-node hub and the ring. If hubs have very similar sizes, the same pattern is detected but the difference between the hubs' associated eigenvalues are small. Thus, in Fig. 4 (b), a small constant difference is denoted by the gaps up to $i = c - 1$.

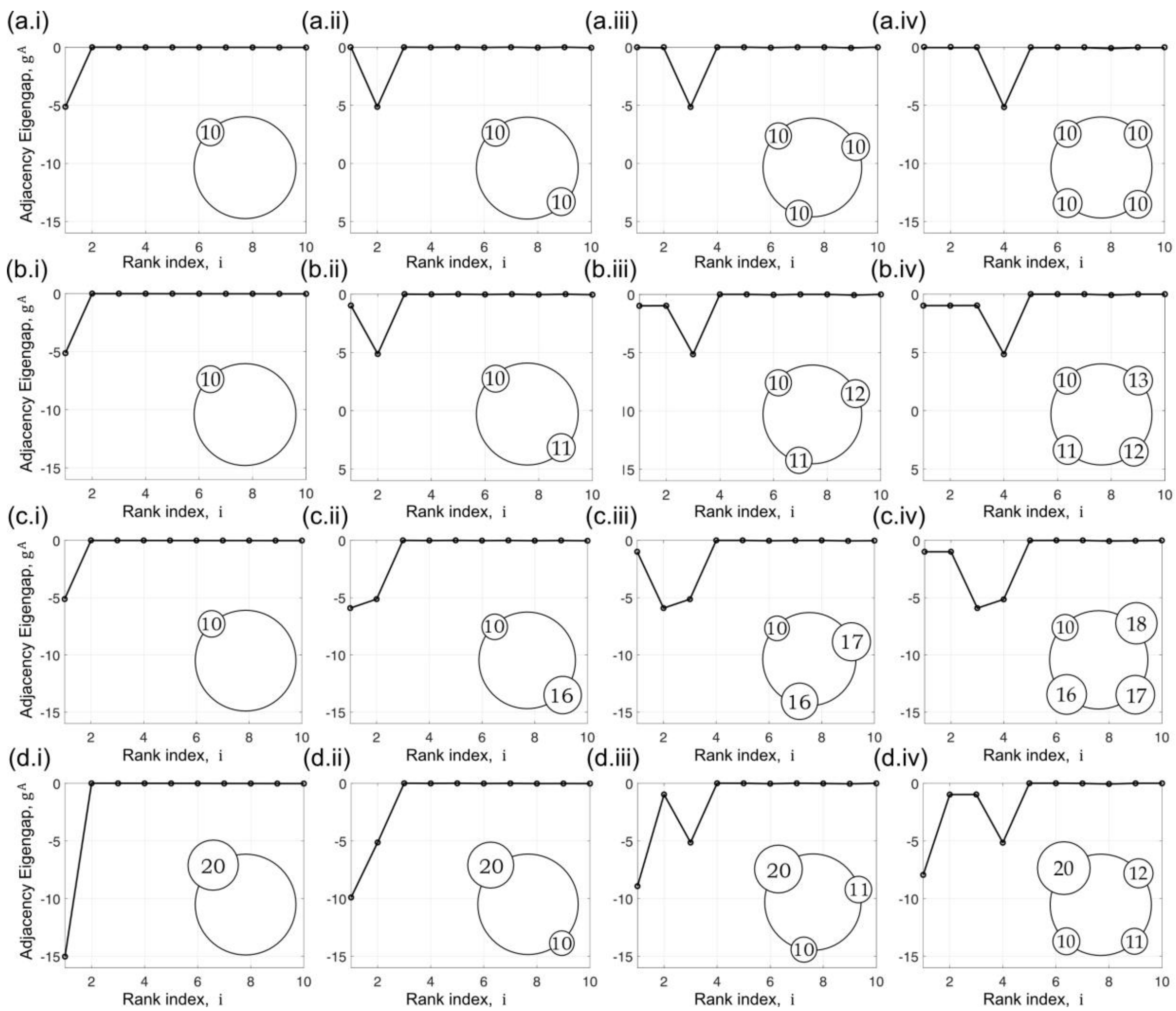

Fig. 4: ***Adjacency*** eigengap spectra of a 200-node ring, connectivity $\delta_R = 4$, embedded with an increasing number of hubs, $c = [1,4]$, of different sizes, (a) $S_1$ (10, 10, 10, 10), (b) $S_2$ (10, 11, 12, 13), (c) $S_3$ (10, 16, 17, 18), and (d) $S_4$ (20, 10, 11, 12). Only key eigengaps (ranks 1 to 10) are shown out of the 200 gaps.

## B. Influence

To study the insights of the Laplacian eigenspectrum, four 200-node aggregated ring networks with connectivity $\delta_R = 4$ are analysed. The simulations are $S_1$ to $S_4$ (Eq. (12) - (15)). Only the cases comprising four hubs, $c = 4$, are examined. The Laplacian eigenvectors observed in Fig. 5 (e) – (h) decompose the salient structures (hubs) highlighted by the adjacency spectra into their component nodes' influence profiles. Nodes with equivalent *influence profiles* group themselves together and are detected by the Laplacian eigenspectra's significant eigengaps (Fig. 5 (a) – (d)). For the ring networks in Fig. 5, all nodes have access to all pathways through the network therefore these influence profiles emerge because of variations in nodes' degree. In other, less regular networks, degree may not necessarily indicate similar influence profiles, e.g. low degree nodes could have access to more nodes through pathways that are unavailable to a higher degree node.

Unlike the adjacency spectra (Fig. 4), the Laplacian identifies a minimum of $|h_k| - 1$ eigenvalues associated with each hub $k$'s nodes. This is observed most prominently in Fig. 5 (d) and (h), where the 19 eigengaps and eigenvector of the 20-node hub appear distinctly at the start of the Laplacian spectrum. The magnitude of Laplacian eigenvalues is related to node degree, with the largest eigenvalues being associated with the most well-connected nodes. This more detailed decomposition of the topology (compared to the adjacency's) allows the Laplacian eigenspectrum to be studied to identify influence patterns which emerge due to degree variations within the hubs. The eigengaps in turn reflect the differences in nodes' influence. If node degrees are particularly differentiated from one another (Fig. 5 (d) and (h)'s 20-node hub), significant differences will be observed between the eigenvalues. Nodes with the same degree and thus influence profiles are detected together by the Laplacian spectra, forming groups of nodes with the same (or similar) influence profiles (see the simultaneous detection of nodes from all four 10-node hubs in Fig. 5 (e)'s eigenvectors).

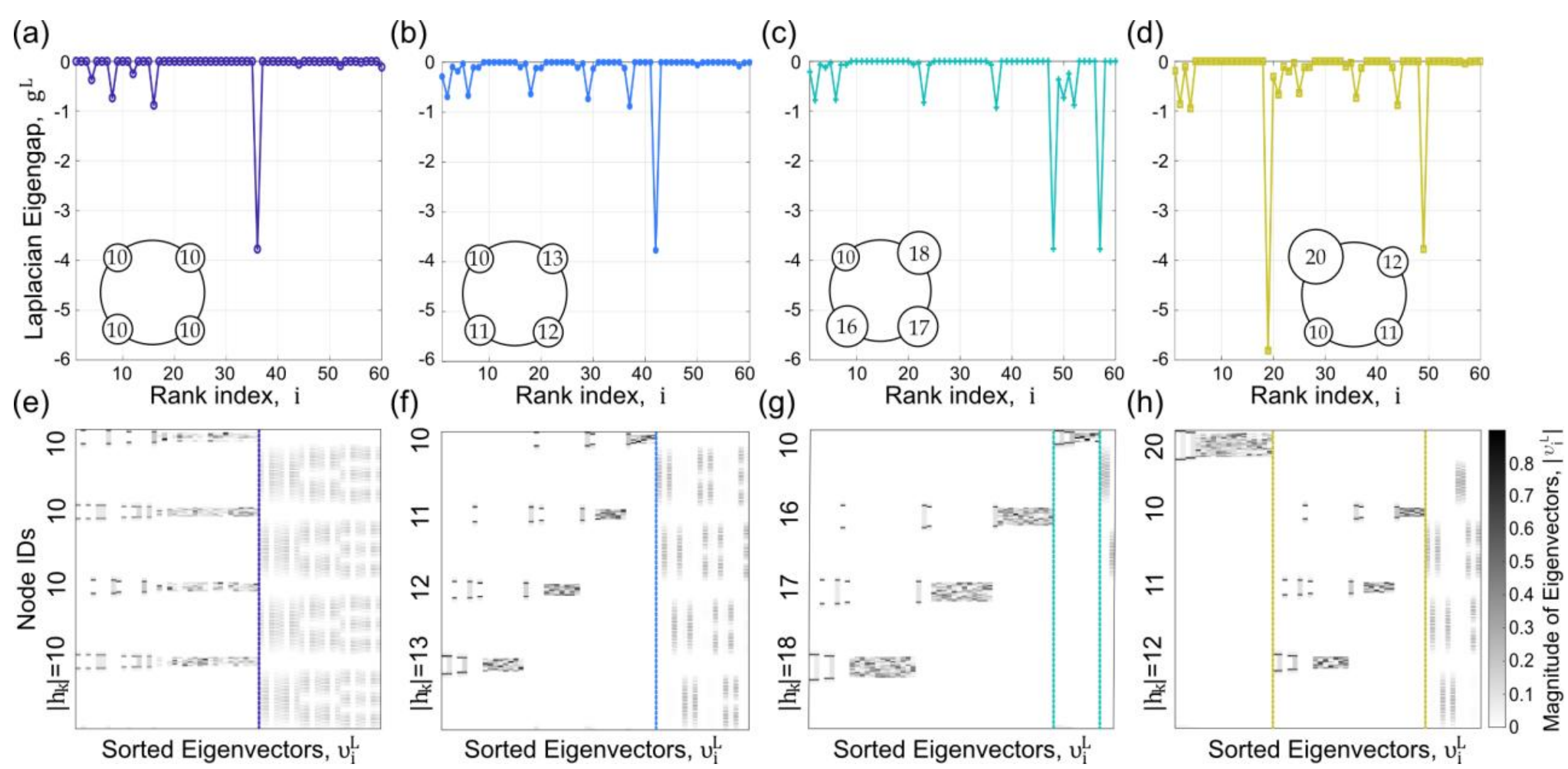

Fig. 5: ***Laplacian*** eigengap spectra and eigenvector's magnitude maps of a ring embedded with four hubs, $c = 4$. The embedded hubs have different sizes, (a)/(e) $S_\alpha$ (10, 10, 10, 10), (b)/(f) $S_1$ (10, 11, 12, 13), (c)/(g) $S_2$ (10, 16, 17, 18), and (d)/(h) $S_3$ (20, 10, 11, 12). The rings have 200 nodes and a connectivity $\delta_R = 4$. The largest and most significant eigengaps are shown by dashed lines on the eigenvector maps.

Each hub produces a distinct pattern of Laplacian eigengaps corresponding to its nodes' influence, and which, for the toy graph examples, allow for a clear differentiation between the hubs' gateway and central nodes.

Gateway nodes are noted in Fig. 5 (e) – (h), to have larger eigenvector entries for the dominant eigenvalues. Note from the alignment of Fig. 5 (a) and Fig. 5 (e), that the gateway nodes and their large eigenvector entries result in the small non-zero eigengaps at the start of the spectrum.

Central nodes (Fig. 2), that is those nodes with no connections outwith the hub, in turn have large eigenvector entries in the eigenvectors associated with the less dominant eigenvalues, indicating their lower influence compared to the gateway nodes. Central nodes all have the same connections (see the hub structure in Fig. 2), and therefore have the same influence profiles, forming a community. As a result, they are identified by the Laplacian spectrum together (see the blocks of neighbouring nodes with high eigenvector entries in Fig. 5 (e) - (h)).

Central and gateway nodes form two nested groups of nodes with similar influence profiles, with the central, less influential nodes, being the core set. Due to their lower connectivity, $\delta_R$, the ring nodes' eigenspectra signature is visible in the less prominent eigenvectors to the right of the eigenvector mappings. Matching the change in node connectivity (from $\delta_k$ to $\delta_R$ ), the transition from hub to ring nodes is marked by a significant spectral gap given for each hub by

$$\begin{aligned} \mathrm{idx}\left(g_k^L\right) &= |h_k| - 1, \\ &= \delta_k. \end{aligned} \tag{16}$$

where $\mathrm{idx}\left(g_k^L\right)$ is the rank of hub $k$'s eigengap with the largest magnitude, $|h_k|$ is the number of nodes belonging to the hub, and $\delta_k$ is the number of direct connections that a node has within its hub (Fig. 2).

The hubs fingerprints appear in the spectra sorted according to their nodes' relative influence and connectivity. Node connectivity, $\delta_k$, is linked to hub size, where the larger the hub, the more nodes it encompasses, and thus the larger the number of direct connections experienced by its nodes. As such, if the hubs have the same size (Fig. 5 (a) and (d)), their nodes' gateway and central have the same connectivity and therefore the same influence profiles. This results in the gaps (Eq. (16)) coinciding with each other, producing a single significant eigengap at

$$\mathrm{idx}(g_{\max}^L) = \sum_{k=1}^{c} (|h|_k - 1), \tag{17}$$

where $c$ is the hub count.

If the hubs have approximately the same size (Fig. 5 (b) and (f)), their influence profiles are similar enough to cause the intermingling of their spectral fingerprints with a single dominant gap developing as per Eq. (17).

If the hubs are sufficiently differentiated in terms of size, as is the case for $S_3$ (Fig. 5 (c) and (g)) and $S_4$ and (Fig. 5 (d) and (h)), so are their influence profiles which are ranked according to their nodes' connectivity. This is observed most clearly in Fig. 5 (g) and (h), where the hubs appear as blocks of nodes with high eigenvector entries, sorted on the basis of the hub size. Note that eigengaps develop at the end of each block of nodes as per Eq. (16). Thus, in Fig. 5 (d), the gateway nodes of the 20-node hub are detected first with gaps at $i = 2$ and $4$ thanks to their connectivity being the highest in the network. The 20-node hub's most significant eigengap, $\mathrm{idx}\left(g_{\mathrm{k}}^{L}\right)$, appears at $i = 19$ as per Eq. (16). The other hubs, due to their similar sizes and connectivity, produce a combined significant eigengaps at $i = 19 + \sum_{k=1}^{3}(|h|_k - 1) = 19 + 30 = 49$ as per Eq. (17).

## C. Case Study

This case study analyses satellite networks comprising two- and three-ring networks (Fig. 6Fig. 6 (a) and (b)). These are modelled from Walker-Delta satellite constellations with ISL ranges of $4000$ km and inclinations of 56 degree, resulting in a ring connectivity of $\delta_R = 4$. The two-ring satellite constellation consists of 200 satellites (100 per ring) at all altitude of 25 000 km while the three-ring network consists of 210 satellites (70 per ring) at all altitude of 10 000 km. The networks are averaged over ten timesteps as per Eq. (11). Note that due to the time averaging, connectivity between nodes takes on non-integer values, making the graph edges weighted.

Building upon the findings of the toy graph analyses, the adjacency spectrum is once again found to provide insights into the relative sizes and number of the networks' salient structures (hubs). In contrast, the Laplacian is found to decompose the salient structures, identifying groups of nodes with similar influence profiles. Although both the adjacency and the Laplacian provide relevant network insights when analysed independently, it is together that they provide the most complete picture of the network, as noted through the analyses of the satellite networks of Fig. 6.

### *1. Structure*

Supporting what was observed in the toy graph examples, the adjacency spectra identify the number of salient structures (hubs) present in the network. In Fig. 6 (c), the spectra's largest eigengaps denote the existence of the two hubs ($c = 2$) created at the two rings' intersection (Fig. 6 (a)), and six hubs ($c = 2$) within the three-ring network (Fig. 6 (b)). The adjacency spectra also confirm what can be hypothesised visually: the hubs have the same size, as denoted by the single non-zero eigengap in both spectra. Should they have been noticeably different, multiple non-zero eigengaps would be expected as was the case for the toy graph example (Fig. 4). Instead, the single non-zero eigengap, compounded by knowledge that the rings have the same focal point (the Earth around which the satellites are orbiting) allow for the assumption that that the hubs are the same. The spectra provide no further information regarding the size and internal structure of the hubs.

### *2. Influence*

The hubs developing at the rings' intersections have the same size (as supported by the single non-zero adjacency eigengaps of Fig. 6), which allows the assumption that the patterns of influence detected by the Laplacian are equally spread across the hubs. This allows Eq. (17), derived from the aggregated rings' analyses, to be used to calculate the number of nodes present in each set of influence profiles.

The Laplacian eigenspectra uncover two nested node groupings within the networks' salient structures (hubs), where the node groups are sets of nodes sets with the same (or equivalent) influence profiles. These are detected by the two dominant eigengaps observed in Fig. 6 (d). The first gap corresponds to the networks' most influential nodes (central nodes), the innermost of the two nested sets. These nodes have the largest eigenvector entries of the dominant eigenvectors (Fig. 6 (e) and (f)) and thus have higher effective leadership over the network. The second significant gap of Fig. 6 (d) coincides with the second nested set which comprise central nodes (the first set) and gateway nodes surrounding them. These gateway nodes have small eigenvector entries for the dominant eigenvalues, but larger ones for the proceeding eigenvectors. As per Eq. (17), within the two-ring network, the first gap at $i = 10$ indicates that six central nodes form the hubs' subset of most influential nodes ( $\mathrm{idx}(g_{\max}^{L}) = \sum_{k=1}^{2}(|h|_k - 1) = 10$ ). The second gap at $i = 18$ indicates the combined detection of central and gateway nodes (10 per hub). For the three-ring scenario, six central nodes are identified in each of the six hubs with the first gap at $i = 30$, followed by the identification of core nodes (central and gateway nodes) with the second gap at $i = 42$ (8 per hub). The other non-zero eigengaps following both networks' largest gap indicate less influential nodes surrounding the core nodes. Their lower influence is noted from them only having large eigenvector entries for the eigenvectors to the right of the spectrum (Fig. 6 (e) and (h)). These gaps account for the nodes surrounding the two nested node sets whose degrees are only slightly higher than the ring connectivity, $\delta_R$.

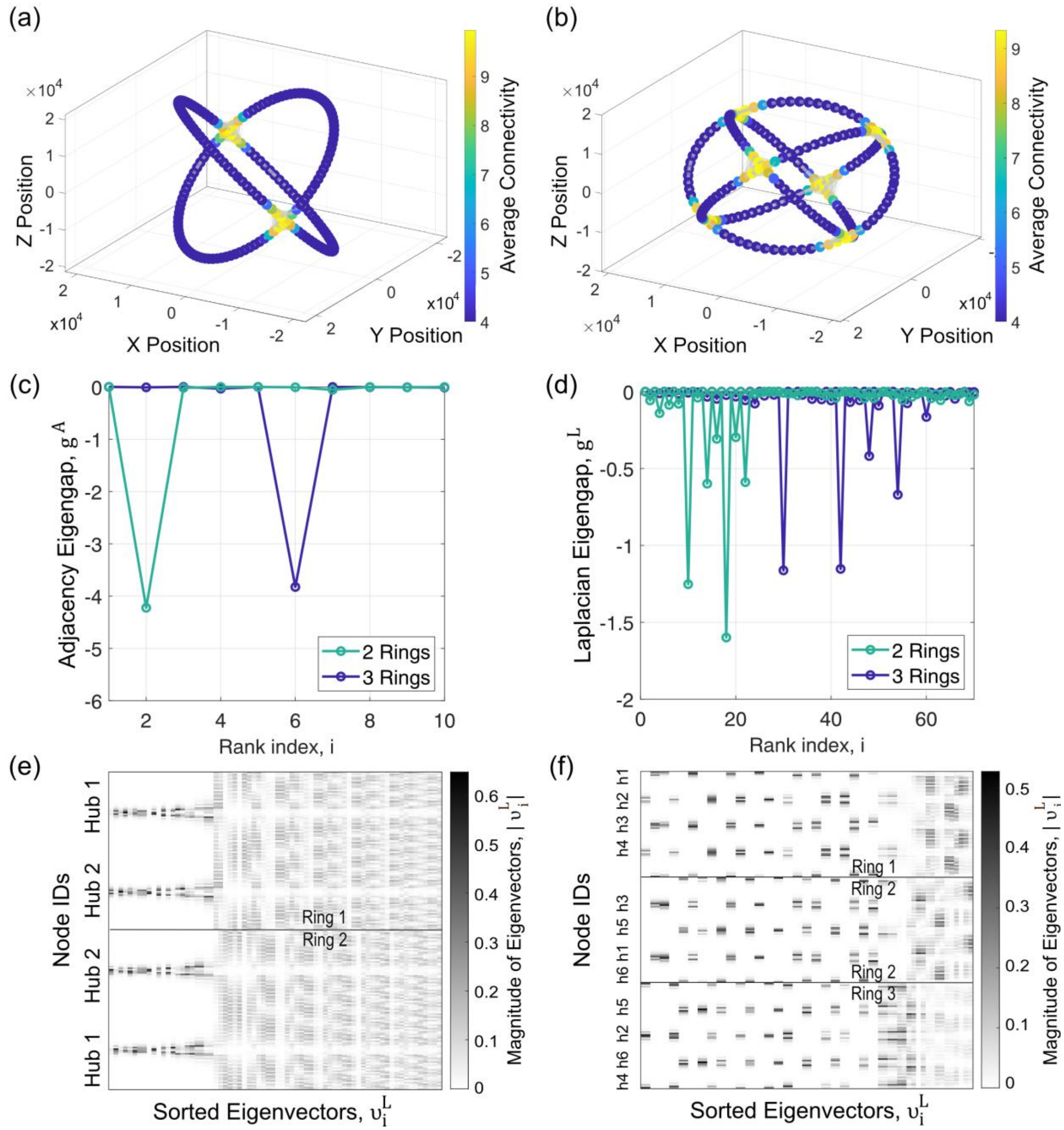

Fig. 6: Intersecting ring networks, their eigengap spectra, and their eigenvectors' magnitude. (a) two rings network with connectivity $\delta = 4$, (b) three rings network with connectivity $\delta = 4$, (c) adjacency eigenspectra, (d) Laplacian eigenspectra, and Laplacian eigenvectors' magnitude mapping of the (e) two- and (f) three-rings networks.

The hubs' two nested sets of influence profiles can also be visualised by assessing the nodes' local eigenvector centrality as per Eq. (4). Focusing on the two rings network, the network's central and gateway nodes can be distinguished (Fig. 7 (a) and (b)), with the hubs' cores highlighted in Fig. 7 (c). It can also be seen in Fig. 7 (a) that the two sets of influence profiles are not actually centred around the rings' intersections, but rather at a point between the rings in the direction of the nodes' (satellites') movement. This stems from the averaging process used to create the satellite networks analysed herein (Eq. (11)), which accounts for the satellites' orbits and thus shifts the centre of the hubs. Another important point to note from Fig. 7 (a) and (b)'s centrality measures is that the sets of central and gateway nodes are not fully disjoint, meaning that there is overlap in their nodes' influence. This is indicative of the fact that hub nodes are on a spectrum of connectivity and influence and as such, the differentiation between the two sets of influence profiles is not as stark and obvious as was the case for the aggregated rings and their artificial hubs (Fig. 5).

# IV. DISCUSSION

System functions are often related to a network's community structure, that is, the patterns in edge density which emerge at a global and local scale [1, 2]. Community structure can be detected by spectral analysis [7-9]. The insights derived from such analyses depend on the choice of representative matrix, typically the adjacency or Laplacian. However, to the best of the authors' knowledge, no clear intuitions have been derived regarding what matrix might provide the most relevant community detection results, but also what complementary insights the analyses of both matrices may offer.

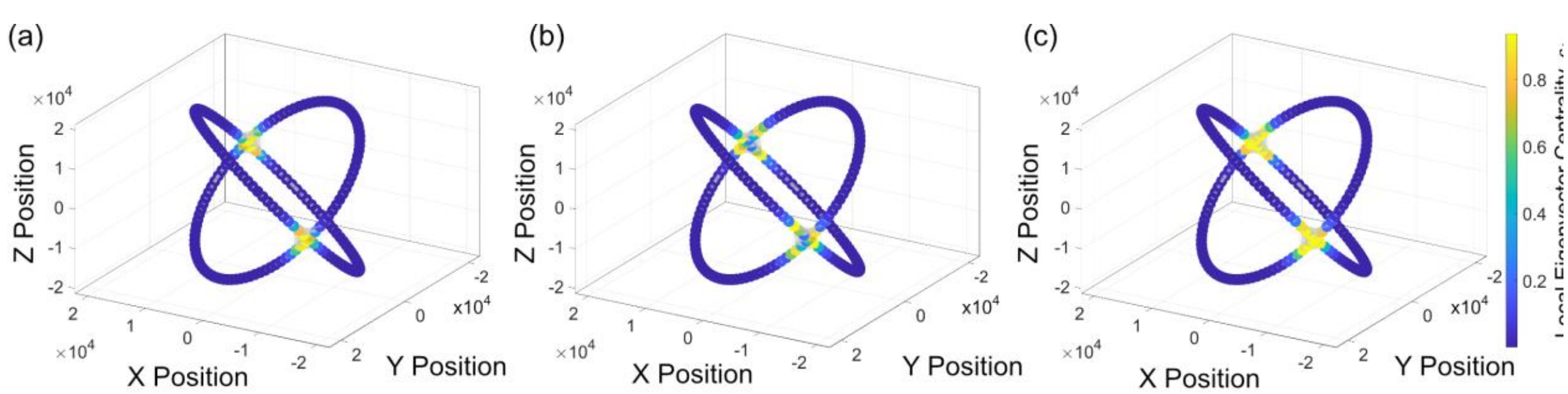

Fig. 7: Laplacian local eigenvector centrality of the two intersecting rings satellite network calculated for the eigenvectors spanning (a) $i = 1$ and $i = 10$ (rank of second most significant eigengap), (b) $i = 10$ and $i = 18$ (rank of most significant eigengap), and (a) $i = 1$ and $i = 18$.

The adjacency eigenspectrum is often used for community detection purposes [6, 15]. In the toy examples of Fig. 4, the adjacency spectrum reflects the graph structure, allowing for the identification of the number and relative sizes of the hubs embedded on the rings. However, the analyses only identify the most salient structures the networks. This is because a single eigenvalue is observed to be associated with each hub, with the eigenvalue's magnitude relating to the hub size. Although changes in the hubs' sizes are apparent, this coarse breakdown of the network topology means that for sparse network such as the one of aggregated rings, the adjacency cannot offer insights into the influence patterns existing within the networks' salient structures.

Other than the adjacency, the Laplacian is often the matrix of choice for community detection purposes due to its associations with consensus dynamics [23-25]. In the study of the toy graph examples of Fig. 5, the Laplacian decomposes the salient structures identified by the adjacency spectra in terms of their influence profiles. This finer analysis is possible because, unlike the adjacency, for a hub of size $|h_k|$, the Laplacian associates $|h_k| - 1$ eigengaps per hub. The Laplacian eigenvalues magnitudes are related to the nodes' degree and thus, their ability to influence the network. The most influential nodes, meaning those having the largest eigenvector entries for the spectrum's dominant eigenvectors, are consistently found to be those most well-connected nodes. The Laplacian spectrum sees nodes grouped in sets with the same or similar influence profiles denoted by significant eigengaps. Given knowledge of the relative sizes of the hubs from the adjacency eigenspectrum, the ranks at which these eigengaps develop can be used to derive the size of the node sets with equivalent influence profiles (Eq. (16)).

The case study of satellite networks (Fig. 6) demonstrates that the adjacency and Laplacian insights gathered from the toy graph examples hold for a model of a real system. It also supports the idea that for sparse networks such as those of satellite constellations, the adjacency and Laplacian spectra of complex networks are best interpreted together. The adjacency provides key information on the number and relative sizes of salient structures but says nothing regarding the number of nodes present in those structures, nor their relative connectivity. In turn, the Laplacian decomposes these structures into influence profiles, forming groupings given the nodes' effective leadership and influence, where influence is highlighted by nodes' eigenvector entries. However, this decomposition requires knowledge of the number and relative sizes of the hub structures to know how the nodes' influence is spread across the network's salient structure. The case study also demonstrates that Laplacian analyses can uncover the hierarchical ordering of nodes' influence profiles, along with their groupings.

## V. CONCLUSION

Comparative analyses of adjacency and Laplacian eigengap spectra of ring-based networks support the established knowledge that both matrices provide community detection insights. In networks of aggregated and intersecting satellite rings, the adjacency eigenspectrum reveals graph structure, identifying the number and relative sizes of salient structures. In contrast, the Laplacian decomposes the salient structures and identifies the influence profiles which exist within them, due to variation in node connectivity, where sets of nodes with similar influence profiles are grouped by Laplacian eigengaps. However, these Laplacian insights must be contextualised by knowledge of the number and relative sizes of the salient structures existing in a network – knowledge obtained from the adjacency analyses. Although independent analyses of the adjacency and the Laplacian both provide insights into network characteristics, for sparse networks such as those of satellite constellations, it is together that they best detect a network's community structure.

# DECLARATIONS

## Data Availability

No data were created or analysed in this study, however the simulation code is available at DOI: 10.5281/zenodo.17661737.

## Acknowledgements

This work is funded by the Air Force Office of Scientific Research (AFOSR), Grant number FA8655-22-1-7033.

## Author Contributions

Conceptualisation and methodology, A.B., R.A.C., and M.M.; numerical investigation A.B.; interpretation, A.B., R.A.C., and M.M.; writing, review, and editing, A.B., R.A.C., and M.M.; funding acquisition, M.M and R.A.C. All authors have read and agreed to the published version of the manuscript.

# APPENDIX

## Appendix A

As defined in Eq. (11), the satellite networks over which the spectral analyses are performed are time-averaged over a window, $tw$, comprising multiple time steps, $\tau$. The choice of the averaging window duration is highlighted in Fig. 8. Therein, the magnitude of relevant Laplacian eigengaps (the first 25 for the two-ring network and the first 50 for the three-ring network) is shown against an increasing averaging window, demonstrating the window's impacts on the insights provided by the spectral analyses. As observed in Fig. 8, for small windows, the instantaneous connections dominate the structure, resulting in noisy spectra where no clear node sets with similar influence profiles are detected – where node groupings are defined by the spectra's dominant eigengaps. In contrast, increasing the averaging window overmuch leads to the hubs being "stretched". As a result, the hubs see lower levels of connectivity and become increasingly homogeneous, losing their community structure. To balance these two effects, an averaging window of ten time steps, $tw = 10$, is selected for the analyses of both networks.

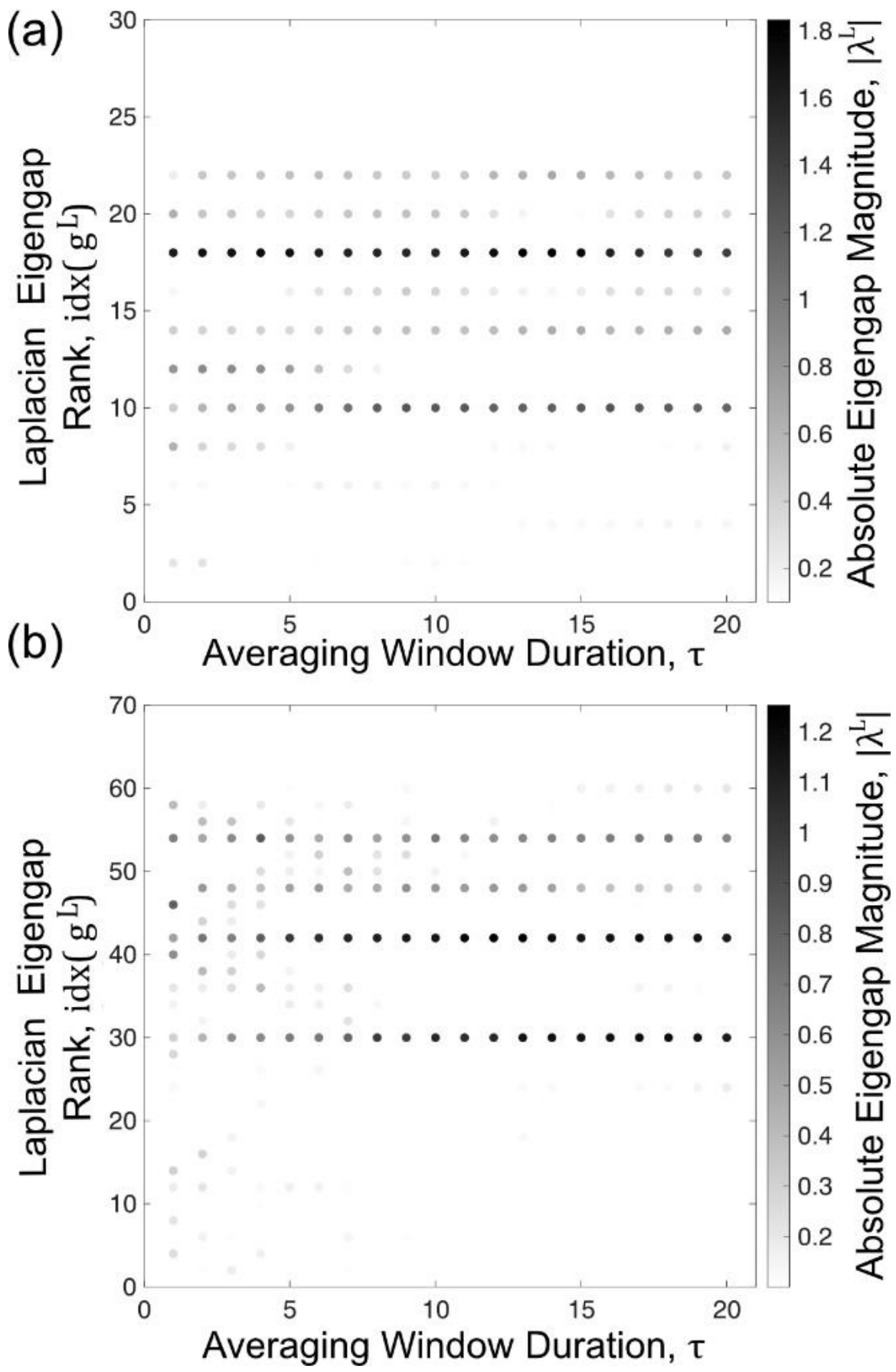

Fig. 8: Impact of the averaging window duration, t, on the Laplacian eigenvalues' magnitudes of the (a) two- and (b) three-ring networks.